\documentclass[aps,prd,preprintnumbers,nofootinbib,twocolumn]{revtex4}
\usepackage{bm}
\usepackage{latexsym}
\usepackage{dcolumn}
\usepackage{amsmath,amsfonts,amssymb}
\usepackage{graphicx,epsfig}
\usepackage{psfrag}
\usepackage{amsthm}

\def\be {\begin{equation}}
\def\ee {\end{equation}}
\def\bea {\begin{eqnarray}}
\def\eea {\end{eqnarray}}
\def\bc {\begin{center}}
\def\ec {\end{center}}
\def\bfg {\begin{figure}}
\def\efg {\end{figure}}
\def\bi {\begin{itemize}}
\def\ei {\end{itemize}}

\def\beq{\begin{equation}}
\def\eeq{\end{equatfion}}
\def\br{\begin{eqnarray}}
\def\er{\end{eqnarray}}
\newcommand{\eel}[1] {\label{#1}\end{equation}}

\newcommand{\bdm}{\begin{displaymath}}
\newcommand{\edm}{\end{displaymath}}

\begin{document}
\title{Brief reply to ``On the correctness of cosmology from quantum potential"
}

\author{Ahmed Farag Ali $^{1,2}$}\email[email: ]{ahmed.ali@fsc.bu.edu.eg; afali@fsu.edu }
\author{Saurya Das $^3$} \email[email: ]{saurya.das@uleth.ca}

\affiliation{$^1$ Department of Physics, Florida State University, Tallahassee, FL 32306,  USA.\\}
\affiliation{$^2$ Dept. of Physics, Faculty of Sciences, Benha University, Benha, 13518, Egypt.\\}

\affiliation{$^3$ Department of Physics and Astronomy,
University of Lethbridge, 4401 University Drive,
Lethbridge, Alberta, Canada T1K 3M4 \\}

\begin{abstract}
We respond below to the comment of E. I. Lashin [arXiv:1505.03070]
on our work Phys. Lett. {\bf B741} (2015) 276-279 [arXiv:1404.3093], and point out the errors in that comment.
\end{abstract}

\maketitle

In this brief note, we point out the elementary errors made by the author in his analysis
\cite{lashin}.

\vspace{0.2cm}
\noindent
{\bf Quantum Raychaudhuri equation}

As clearly mentioned at the beginning of \cite{qre}, a fixed classical background was assumed throughout the paper, meaning an associated classical background metric {\it and} the induced metric, which do not depend on
$\hbar$. 
These are then used to derive the Quantum Raychaudhuri equation (QRE).
The author of \cite{lashin} define the so-called `projector' as an alternative to our induced metric
in terms of the quantum velocity field, which has no meaning in the $\hbar \rightarrow 0$ limit
(it would have a $\hbar$ either in the definition of the eikonal,
or the quantum velocity field), and therefore the rest of their analysis using this
projector is also inconsistent
(the quantum corrected Newton's law, geodesic equation, QRE etc. on the other hand smoothly go over to
their classical counterparts in the $\hbar \rightarrow 0$ limit, as they ought to).
Furthermore, as stated more than once in \cite{qre}
(see also refs.[3,7,8] therein), quantal (Bohmian) trajectories do not
meet or cross, a result which follows from the properties of first order differential equations governing these trajectories, without the need for any further assumptions
(e.g. this property holds
for spacetimes with or without symmetries and trajectories with or without shear or torsion).
This alone ensures that there are no conjugate points for quantal trajectories,
rendering the singularity theorems inapplicable. 
This was the main result of \cite{qre}.
Also as the author may be aware, the Raychaudhuri equation does incorporate dynamics,
and important conclusions can be drawn from it, when
curvature terms are replaced using the Einstein equations.
This extends to quantum dynamics when quantal trajectories are used.

\vspace{0.2cm}
\noindent
{\bf Quantum Friedmann equation}

Contrary to the claim of the author of \cite{lashin}, the formalism and results
of \cite{qre,alidas} are covariant, as can be seen for example from Eqs.(17,18) of \cite{qre}
and Eq.(2) of \cite{alidas}(the quantum Friedmann equation or QFE).
Application of a manifestly covariant formalism to specific solutions of Einstein equations
written in certain coordinates, for example to get Eqs.(7-11) of \cite{alidas},
do not break covariance.
Once again, the infinite age of our Universe obtained in \cite{alidas} is simply a manifestation of
the no-crossing property of Bohmian trajectories, in this case to the evolution of
various points of the cosmic fluid, alluded to earlier.
Also as explained in \cite{alidas}, and later in more detail in
\cite{dasbhaduri}, the wavefunctions with a large spread used are fairly accurate, although
not exact descriptions of a
homogeneous and isotropic universe, and for which one rigorously obtains a small and constant
$\Lambda_Q$. Of course there will be additional higher order corrections, and possibly also corrections from
fluctuations of spacetime itself, if a consistent way to compute such systematic
corrections are found.

Many of its intermediate steps of \cite{lashin}
are also incorrect.
E.g. Eq.(28)of \cite{lashin}, in that it uses the $4$-dimensional metric,
as opposed to the induced $3$-dimensional metric in contractions;
Eq.(29), although quite close to Eq.(7) of \cite{alidas},
is erroneous because to derive this, the author uses the classical $H^2=8\pi G\rho/3$, whereas the
integrated QRE, in conjunction with the continuity equation should have been used instead.
Alternatively, its correct version would follow from a simple substitution $H=\dot a/a$ in the
QFE, instead of the several roundabout steps suggested in \cite{lashin};
consequently their statements following Eq.(29), related to values of $\omega$ and fixed points
in the past, which are meaningless as well.
But since these details are mostly irrelevant to the discussions
of \cite{qre,alidas}, there is no need to elaborate.

To conclude, the main results of \cite{qre,alidas} follow from the well-known no-crossing
property of quantal (trajectories), which \cite{lashin} seems to have missed entirely. 
In those papers, we also
developed the formalisms suited for the specific problems at hand, namely
the singularity theorems for quantal trajectories and the QFE.
The author of \cite{lashin} starts with some basic misunderstandings of our work,
as well as of the dynamics of trajectories in spacetimes obeying general relativity.
The author does correctly point out however
that the quantal (Bohmian) trajectories are not guaranteed
to remain timelike throughout. But unlike classical trajectories,
position, momentum or any other observable are not measurable for quantal trajectories 
at any intermediate point either; the measurable predictions in the end are all correct and verifiable.
This is analogous to the non-conservation of energy-momentum for virtual
particles, which are not measurable either, yet consistent with the quantum uncertainty
principle. Here too, all properties associated with the quantal trajectories
are consistent with the uncertainty principl, the standard formalism of
quantum mechanics and causality is not violated (see e.g. refs.[3,5] of \cite{qre}).
%
%
Any sequel from the author of \cite{lashin}
based on a similar set of flawed assumptions should easily be refutable by the careful reader,
by comparing with our original analyses, and
other established body of literature on quantal trajectories, Bohmian mechanics and
the Raychaudhuri equation.
%



\end{document}